# Gamers Private Network Performance Forecasting

From Raw Data to the Data Warehouse with Machine Learning

and Neural Nets

Albert Wong
Mathematics and Statistics,
Langara College
Vancouver BC, Canada
alwong@langara.ca

Chun Yin Chiu
Data Analytics
Langara College
Vancouver BC, Canada
edwardchiu.in.ca@gmail.com

Gaétan Hains
LACL
Université Paris-Est
Créteil, France
gaetan.hains@gmail.com

Jack Humphrey
Computer Science
Okanagan College
Kelowna BC, Canada
jackrbhumphrey@gmail.com

Hans Fuhrmann
Computer Science
UBCO
Kelowna BC, Canada
hansfuhrmann870@gmail.com

Youry Khmelevsky
Computer Science
Okanagan College
Kelowna BC, Canada
ykhmelevsky@okanagan.bc.ca

Chris Mazur
WTFast
Kelowna BC, Canada
christopheramazur@gmail.com

## ABSTRACT

Gamers Private Network (GPN®) is a client/server technology that guarantees a connection for online video games that is more reliable and lower latency than a standard internet connection. Users of the GPN® technology benefit from a stable and high-quality gaming experience for online games, which are hosted and played across the world. After transforming a massive volume of raw networking data collected by WTFast, we have structured the cleaned data into a special-purpose data warehouse and completed extensive analysis using machine learning and neural nets technologies, and business intelligence tools. These analyses demonstrate the ability to predict and quantify changes in the network, and demonstrate the benefits gained from the use of a GPN® for users when connected to an online game session.

## 1 Introduction

The Gamer's Private Network (GPN®) technology of WTFast offers an improvement to the standard internet connection, reducing latency and jitter between the client and game server for connections that occur worldwide. Our research team has studied many aspects of the technology, transitioning now from laboratory setups to using massive datasets of real-time networking data. The measurements used provide details of every gamer-server connection, such as the game being played, IP, player and server geolocation, routing hops, latency, and time of day. Previous publications have shown how our data analysis confirmed the superiority of the GPN® over a standard internet connection, and how models can relate latency, the principal measure for quality of experience, to all the other measured variables. In this paper we have the following contributions: (1) we show the design and implementation of a data warehouse for processing, storing, summarizing, and visualizing such massive real-time data, as well as our (2) latest machine-learning models to predict network latency (measured in ping values) using various environment, network, volume, and game-related variables; (3) our experiments point to the possibility of developing machine learning models that can accurately predict ping values that can further improve the GPN® performance. Other latency-related variables of interest, such as flux and loss, could also be modelled similarly in the future. Neural net models also were investigated, but the machine learning models demonstrated better results. Finally, (4) we present how data analyses could relate to the *genres* of games being played to relate networking performance with game quality of experience.

## 2 Network Latency and Online Games

Featured in our paper is a focus on the implications that genre has on the performance and user satisfaction of the GPN®. Through the continued work with latency prediction modeling, performing a cost-benefit analysis to determine the need for improved playability of a game based on the genre can provide an important metric for determining how to move forward with improving the service. Removing latency in a way that is meaningful for users is the core challenge for WTFast, and the implementation of genre analysis is a useful step forward in that task. Our work here expands upon previous work we have accomplished with regards to genre [28].

The primary genres being considered for online video games in this paper are the following [2, 26]:

- First Person Shooter (FPS)
- Massively Multiplayer Online Role-Playing Games (MMORPG)
- Real-Time Strategy (RTS)
- Multiplayer Online Battle Arena (MOBA)
- Sports Games

In papers chronicling how latency affects online games, Quality of Experience was determined through surveys to people who played games of various genres. In a paper that surveyed users of many different FPS games [26], a latency of 80 milliseconds was determined to be an acceptable latency, whereas a paper surveying user from several MMORPG's [27], a latency as high as 120 milliseconds was considered excellent.

Using this information, we will summarize our data warehouse for managing networking data for video game players and servers hosted on the GPN®. We then present the conclusions of our in-depth study of machine-learning techniques for analyzing and possibly predicting network parameters, specifically latency as influenced by all other parameters of the online gaming session. Finally, we will present our first steps in the direction of a rational study of game genres as they relate to latency, with the goal of defining a useful and objective quality of experience of GPN® users.

## 3 Raw Data and Data-Warehouse

### 3.1 Data Columns and Measures

WTFast's GPN® operations generate anonymous performance data that our project is using to analyze and understand the complex real-time interactions between players, games, servers, and networks. The raw data is organized with data analysis and visualization in mind. Relevant data fields include the following ('geo' means "geolocation." Geographical coordinates are a critical parameter for GPN® or internet routing):

| Table | Measures |
| --- | --- |
| Client | Client_IP, Client_ISP, Reg_Country, Client_Geo |
| Server | Server_ID, Game_IP, Game_ISP, Game_Name, Game_Geo |
| Calendar | Session_Start_Day, Session_Start_Month, Session_Start_Year, Session_Start_Hour, Session_Start_Minute, Session_Start_Weekend, Session_End_Day, Session_End_Month, Session_End_Hour, Session_End_Minute, Session_End_Weekend |
| Session | Internet_Flux, Internet_Loss, Internet_Spke, Internet_Ping, WTFast_Flux, WTFast_Loss, WTFast_Spke, WTFast_Ping, Bytes_Total |

**Figure 1: Data Warehouse Dimensions and measures**

A wealth of information is being gradually extracted from this data and it should prove beneficial. The main correlation of interest we look at is the information captured between the session performance and the other three dimensions.

### 3.2 Business Analytics

We created a data visualization and querying dashboard so that a GPN® operator or analyst can access detailed and complex properties of the data. Views were based on game players, routers, game servers, and individual games. We demonstrated the effectiveness of the GPN® while also being able to isolate potential customer markets that may feel dissuaded from a GPN® due to underwhelming performance gains.

The focus was to visualize the information within the data warehouse to try find individual and combined attributes that could be used to determine what factors caused lower-than-anticipated network improvements. Looking at the data through a Business Intelligence tool helped reinforce our understanding of what attributes impacted performance, and by how much. Geographically, the impact that time of day, and game genre by region were the largest outliers that helped guide future decisions.

## 4 Machine-Learning Models for GPN® Data

In this section, we will describe the processes employed to model network latency using ping values under WTFast GPN® routing. The individual play session and client records collected by WTFast over the summer and fall of 2020 was stored in our data warehouse and was used as the data source.

Before loading data into the data warehouse, we used R and a bash script for the cleaning and merging process. Within Figure 2 is the list of data prior to the transformation and loading into the data warehouse.

| Name | No. of files | Data Entries |
| --- | --- | --- |
| July Plaid Data | 1 | 369,479 |

| Name | No. of files | Data Entries |
|---|---|---|
| August Plaid Data | 61 | 357,151 |
| September Plaid Data | 61 | 330,304 |
| September Client Stats Data | 43,200 | 3,659,973 |
| October Plaid Data | 62 | 339,534 |
| October Client Stats Data | 44,638 | 3,557,641 |

**Figure 2: Summer and Fall 2020 data used in data warehouse**

## 4.1 Pre-Processing and Data Transformation

We applied several rules to the play session records to ensure that the ping value we used, WTFast_Ping, has meaningful values (non-null, etc.) We also only included records where the game duration and a "calculated distance" variable that could be calculated using longitude and latitude values for the client and game server.

| Numerical Features | Description |
|---|---|
| INTERNET_PING | Average internet ping in milliseconds for the game session |
| INTERNET_FLUX | Average internet flux in milliseconds for the game session |
| INTERNET_LOSS | Cumulative count of the Internet packets lost for the entire game session |
| INTERNET_SPKE | Cumulative count of internet spikes for the entire game session |
| BYTES_UP_TCP | Total bytes sent from client to game server via TCP connections |
| BYTES_UP_UDP | Total bytes sent from client to game via UDP connections |
| BYTES_DOWN_TCP | Total bytes received by the client from the game server via TCP connections |
| BYTES_DOWN_UDP | Total bytes received by the client from the game server via UDP connections |
| SOCKET_COUNT_TCP | Count of TCP socket connections made |
| SOCKET_COUNT_UDP | Count of UDP socket connections made |
| CLIENT_IP_COUNT | Count of unique client IP addresses, excluding proxy IP addresses |
| GAME_IP_COUNT | Count of unique game server IP addresses, excludes proxy IP addresses |
| BYTES_PER_SECOND | Average bytes per second from the game session |
| CALCULATED_DISTANCE | Total distance in Megameters from client to the node and from node to game |
| DURATION | Total duration in seconds from first socket connection open to final socket connection close |

**Figure 3: Description of all numerical features in the data set**

We began with the above 14 numerical features in the modelling process. They included latency variables if the game session was routed through a standard internet connection or the GPN®, volume variables such as Bytes_Up or Bytes_Down, as well as variables related to the configuration of the network used.

| | Percentile | | | | |
|---|---|---|---|---|---|
| **Numerical Features** | **25** | **75** | **85** | **95** | **Max** |
| INTERNET_PING | 65 | 211 | 256 | 325 | 500 |
| INTERNET_FLUX | 1 | 7 | 12 | 27 | 368 |
| INTERNET_LOSS | 0 | 12 | 32 | 162 | 25770 |
| INTERNET_SPKE | 0 | 30 | 72.6 | 255 | 13217 |
| BYTES_UP_TCP | 0.02 | 2.47 | 4.98 | 12.31 | 564.01 |
| BYTES_UP_UDP | 0 | 6.25 | 16.37 | 52.54 | 32226.17 |
| BYTES_DOWN_TCP | 0.06 | 26.54 | 52.14 | 136.6 | 63300.35 |
| BYTES_DOWN_UDP | 0 | 22.70 | 58.89 | 181.1 | 30650.83 |
| SOCKET_COUNT_TCP | 2 | 23 | 34 | 184 | 25405 |
| SOCKET_COUNT_UDP | 0 | 8 | 21 | 90 | 8449 |
| CLIENT_IP_COUNT | 1 | 1 | 1 | 1 | 17 |
| GAME_IP_COUNT | 2 | 8 | 15 | 45 | 2835 |
| BYTES_PER_SECOND | 1.59 | 9.48 | 12.13 | 32.74 | 11316.87 |
| CALCULATED_DIST | 3.05 | 12.15 | 14.93 | 19.84 | 35.48 |
| DURATION | 2623 | 15966 | 26901.4 | 43941.4 | 662624 |

**Figure 4: An analysis of feature values by percentiles**

It can be observed from the table above that the features have very long right-tailed distributions as the duration of some game sessions is very long (the maximum was 662624 seconds, or 7.6 days). A square root transformation was applied to these numerical features to mitigate this potential problem in the modelling process. A min-max normalization was then applied to standardize these features before they were implemented into the machine learning algorithms. The minimum and maximum values for each feature were calculated from the July PLAID data.

48.2% of the game sessions occurred on the weekend (Friday, Saturday, or Sunday). A "one-up" numerical feature was created to evaluate the potential impact to latency of games played on the weekend. Additionally, using an initial classification of the games played, variables for the one-up features were created for the following game types.

The one-up variables were added to the list of numerical features to further evaluate the possible relationship between game type and network latency.

| One-up Variables |
|---|
| WEEKEND |
| RPG.MMP |
| RPG.CASUAL |
| OTHER |
| ACT.SHOOTER |
| ACT.RPG.STRATEGY |
| ACT.MMP.SHOOTER |
| ACT.RPG.MMP.ADV |
| ACT.STRATEGY |
| ACT |
| RPG.SHOOTER |
| STRATEGY |
| ACT.RPG.MMP |
| ACT.RPG.STRATEGY.MMP.SIM.ADV |
| ACT.STRATEGY.MMP.SHOOTER |
| SPORTS |
| ACT.MMP.SHOOTER.SIM |
| ACT.MMP |
| ACT.SPORTS |
| ACT.RPG.MMP.ADV.SPORTS |
| ACT.RPG |
| SHOOTER |
| RPG |
| ACT.STRATEGY.MMP |

**Figure 5: List of additional one-up variables utilized**

## 4.2 Machine-Learning Models

Sixty percent of the 55,517 records obtained from the processes described above were randomly selected and used as the training data set for the development of the models. The remaining 40% served as the testing data set and used for evaluating the performance of the models.

A stepwise regression model was built to serve as the baseline model and provide us with a list of features (the "select" list) that were considered statistically significant. This list would be used in the feature engineering process when fine-tuning the models. Using Python and existing libraries (pandas, numpy, math, sklearn, keras, and tensorflow), Multiple Layer Preceptor (MLP), Random Forest (RF), and Support Vector Regression (SVR) models were built using various hyper-parameters. These models were then evaluated using the Root Mean Square Error (RMSE) and the Mean Absolute Percentage Error (MAPE) as the performance metrics.

We also used the technique of binning to deal with the skewness of the numerical features. The features under binning were categorized into groups and "one-up" variables were used to represent the resulting categories. We believe that this approach could improve a "tree-based" algorithm like RF but could penalize algorithms such as MLP or SVR. Our results confirmed that this was correct.

Another feature engineering approach employed in the modeling process was to use the "select" list as input. Results using this approach were mixed.

It is important to mention that RMSE is an unit-sensitive performance metric and should not be used to compare the models with different measurement units (such as WTFast_Ping as the dependent variable versus the log of WTFast_Ping as the dependent variable.) On the other hand, MAPE measures the average percentage of error between the actual and the predicted values of the dependent variable. The only time where this metric does not make sense is when the dependent variable takes on zero values. Since the WTFast_Ping values are greater than 0 in our data set, we can use MAPE as the main performance metric.

| Algorithm | Feature | Model Details | RMSE | MAPE |
|---|---|---|---|---|

| Stepwise Regression | All | Backward Elimination | 37.06 | 68% |
| --- | --- | --- | --- | --- |
| Random Forest | Select List and Selective Binning | n_estimators=400 | 27.47 | 23.0% |
| Random Forest | All | n_estimators=400 | 27.45 | 23.3% |
| Multiple Layer Preceptor | Select List | 4 Hidden Layer (100 - 75 - 50 - 25), activation (relu), optimizer='nadam', loss='mse', epoch = 100, batch_size = 25 | 29.84 | 26.4% |
| Support Vector Regression | All | kernel = 'rbf'; gamma = 'scale'; c = 5 | 35.04 | 38.2% |

**Figure 6: Results of the "best" models according to MAPE**

The distribution of WTFAST_PING was right-skewed and was not normally distributed. Therefore, we used a logarithmic transformation as a way to improve model performance. Figure 6 shows the incremental improvements in RF and MLP models.

| Algorithm | Features | Model Details | RMSE | MAPE |
| --- | --- | --- | --- | --- |
| Stepwise Regression | All | Backward Elimination | 0.42 | 7.1% |
| Random Forest | Select List & Selective Binning | n_estimators=400 | 0.307 | 4.45% |
| Random Forest | All & Selective Binning | n_estimators=400 | 0.308 | 4.5% |
| Multiple Layer Preceptor | Select list | 6 Hidden Layer (150 - 125 - 100 - 75 - 50 - 25 ); activation (relu) optimizer='Nadam'; loss='mse'; epoch = 100; batch_size = 25 | 0.338 | 5.15% |
| Support Vector Regression | All | kernel = 'rbf'; gamma = 'scale'; c = 5 | 0.364 | 5.45% |

**Figure 7: Results of the "best" models for Natural Logarithm**

Following these experiments, we can choose the most effective machine learning models. We also understand that a classification or prediction of latency is best done with a logarithmic transformation. Our research leads us to believe that categories like 0-5ms, 5-50ms, and 50-200ms are what matters most for quality of service for gamers (i.e. human reflex time-scales). It is not necessary for delay categories to be uniform or overly granular. The data analysis is currently offline, but one goal of our project is to use it for large-scale simulations or to utilize real-time network understanding or routing.

## 5 From Quality of Service to Quality of Experience: Game Genres

### 5.1 Game Genres and Types

A video game is said to belong to specific "genres" according to their dominant gameplay features. Gameplay does not refer to the narrative context of the game, but rather the intrinsic qualities and rules of the game, as well as the way the player is expected to interact with the game [24]. Our GPN® data lists the name of the game being played in each session, allowing us to classify them by genre and quantify whether that session experienced low and reliable latency for the corresponding genre. Video game databases such as RAWG (rawg.io) use approximately 20 game genres to characterize the majority of individual games. We have classified more than 1000 of our game sessions for genres and found that the following genres are the most relevant.

| Genre | Game View | Goal |
| --- | --- | --- |
| ACTION | First & Third person | Compete in various competitions. A very large set of sub-genres. |
| RPG | | Complete quests |
| STRATEGY | Third person | Control of map for one army. |
| MMP | | Massively multiplayer game. |
| SHOOTER | First person | Shooting targets. |
| SIMULATION | | Realistic simulation of a given activity while avoiding its real-life disadvantages or risks. |

| Genre | Property | Property | Property |
| --- | --- | --- | --- |
| ACTION | Speed | Skill | Dexterity |
| RPG | Inhabit a character. | Face a sequence of challenges. | Earn new capacities such as powers, weapons. |
| STRATEGY | Armies | Resources | Select the best actions to destroy enemies. |
| SHOOTER | Also exists as "objective shooting" | Considered sub-genre of Action by | |

| | | certain authors. | |
|---|---|---|---|
| SIMULATION | Managing a pet, raising animals, driving a vehicle, managing, etc. | Realism. | Similar to serious games, learning a skill. |

**Figure 8: Breaking down the properties of our primary genres**

As with films or novels, games are usually given more than one genre. For this reason, we have categorized the different combinations encountered into what we have called types. Using this way of viewing each game, ACTION-RPG is one type and ACTION-RPG-MMP is another. This allows us to make clear distinctions between games with similar sets of genres, as in the example given, one may have a few concurrent players, whereas the one featuring MMP could have millions of concurrent players across a game's servers.

Using subsets of the 6 genres for classification can lead to combinatorial explosion and relatively useless distinctions between categories (types) that would gather too few examples in the datasets. Fortunately, the rawg.io database classification of our game sessions produce only 14 types as seen in our lattice diagram. The numbers given for each unique subset of genres is the number of "popular" games belonging exactly to those genres. Popularity is measured by the number of connections in our database where that game is being played. We can see that the distribution of genre sets is relatively balanced so that none of them is, a priori, considered useless.

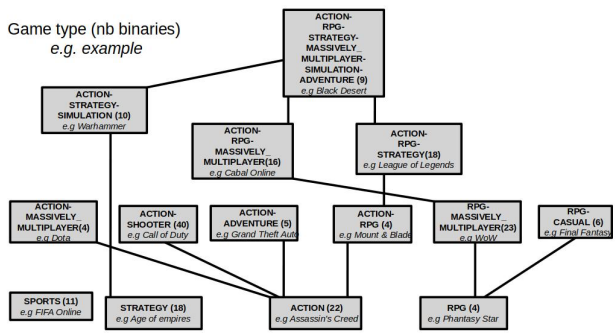

**Figure 9: Lattice diagram showing all genre subsets found**

### 5.2 Relating game Types to Network Performance

There are many relevant and open questions that can be analyzed with the added dimension of game genre/type such as:

- What fraction of action games do players benefit from a real improvement due to GPN®?
- Is the time of day, day of the week, or any date-related category relevant for understanding quality of service (latency) and its perception as the quality of experience?

We have begun work in this direction and the first hurdle is that game types are not uniformly represented in the datasets. This is clear from the parenthesized counts in Figure 9 and the percentage of records seen in these genre subsets in Figure 10. Certain types appear to be very rare for GPN® users, for external human reasons, or purely mathematical facts like having too many (up to 6) genres are definition.

| Game Types | Percentage of Records |
|---|---|
| RPG.MMP | 27.1% |
| RPG.CASUAL | 20.1% |
| OTHER | 11.9% |
| ACT.SHOOTER | 11.6% |
| ACT.RPG.STRATEGY | 10.6% |
| ACT.MMP.SHOOTER | 5.1% |
| ACT.RPG.MMP.ADV | 1.7% |
| ACT.STRATEGY | 1.6% |
| ACT | 1.5% |
| RPG.SHOOTER | 1.3% |
| STRATEGY | 0.8% |
| ACT.RPG.MMP | 0.8% |
| ACT.RPG.STRATEGY.MMP.SIM.ADV | 0.8% |
| ACT.STRATEGY.MMP.SHOOTER | 0.7% |
| SPORTS | 0.7% |
| ACT.MMP.SHOOTER.SIM | 0.7% |
| ACT.MMP | 0.7% |
| ACT.SPORTS | 0.6% |
| ACT.RPG.MMP.ADV.SPORTS | 0.5% |
| ACT.RPG | 0.5% |
| SHOOTER | 0.4% |
| RPG | 0.2% |
| ACT.STRATEGY.MMP | 0.2% |

**Figure 10: A breakdown of existing genres by occurrence**

A purely optimal and balanced partition of the types of lattice would not necessarily make sense for game players, so we refrain from using this for now.

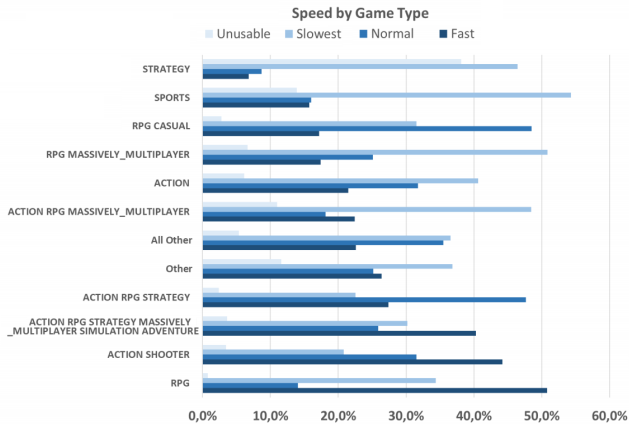

Figure 11: Breakdown of game speeds observed by genre

We have begun relating the genres to their network requirements by assigning each one a value between 1 and 5 for their required networking speed, with 1 being a low, stringent network speed, and 5 being a higher, and comparatively lax network speed. This sensitivity index has been intuitively defined and then adjusted to experimental observations. Once it is given, we proceeded to define a real quality of experience index for GPN® gaming sessions.

| QoS | Quality of Service | An objective experimental parameter inversely proportional to latency and to latency variance |
|---|---|---|
| SI | Sensitivity Index | For a given game, the assigned value between 1 and 5 that indicates minimum network requirements |
| QoE | Quality of Experience | A value determined through the product of Quality of Service multiplied by the Sensitivity Index |

Figure 12: Defined terminology and formulas for quality

Quality of Experience values can be assigned to game types for a whole dataset, to game sessions, to GPN® users, etc. We hope that our evolving data-technologies will allow such quantities to be turned into a deep understanding of the interaction between video gameplay and global networking.

## VI. Conclusions

Internet routing technology that targets reliably low latency enables real improvements in quality of service for online games. It has been shown, through the work described above, that the use of a data warehouse allows for the accumulated storage of formatted and cleaned game session data and facilitates, once it is fully built, the almost real time monitoring and analyses of problems and bottlenecks through data visualizations. On the other hand, the machine learning exercises just completed demonstrate the possibilities of using this data to model and predict network latency measures of game sessions by physical distance, internet environment, game type, and other operational variables. The initial investigation of game genres versus quality of service and quality of experience shows some promise, especially if the sensitivity index could be developed and verified empirically rather than determined subjectively as currently constructed. These three areas of development and enquiry will further the understanding and improvements of online game playability and the requisite network management.

## ACKNOWLEDGMENTS

The research described here is part of the GPNPerf2 research project between Okanagan College and the WTFast company of Kelowna (BC), funded by NSERC CCI ARD Level 2 grant "GPNPerf2" 2019-2021.